\documentclass{INTERSPEECH2023}
\usepackage{url}
\usepackage{booktabs,subfigure,float,amsmath,amssymb}
\usepackage{adjustbox}
\usepackage{threeparttable}
\usepackage[para]{footmisc}
\usepackage{lipsum}

\interspeechcameraready

\title{Generalizable Zero-Shot Speaker Adaptive Speech Synthesis with Disentangled Representations}
\name{Wenbin Wang$^1$, Yang Song$^1$, Sanjay Jha$^1$}
\address{
  $^1$School of Computer Science and Engineering, University of New South Wales, Australia}
\email{wenbin.wang@unsw.edu.au, yang.song1@unsw.edu.au, sanjay.jha@unsw.edu.au}

\begin{document}

\maketitle
 
\begin{abstract}
While most research into speech synthesis has focused on synthesizing high-quality speech for in-dataset speakers, an equally essential yet unsolved problem is synthesizing speech for unseen speakers who are out-of-dataset with limited reference data, i.e., speaker adaptive speech synthesis. Many studies have proposed zero-shot speaker adaptive text-to-speech and voice conversion approaches aimed at this task. However, most current approaches suffer from the degradation of naturalness and speaker similarity when synthesizing speech for unseen speakers (i.e., speakers not in the training dataset) due to the poor generalizability of the model in out-of-distribution data. To address this problem, we propose GZS-TV, a generalizable zero-shot speaker adaptive text-to-speech and voice conversion model. GZS-TV introduces disentangled representation learning for both speaker embedding extraction and timbre transformation to improve model generalization and leverages the representation learning capability of the variational autoencoder to enhance the speaker encoder. Our experiments demonstrate that GZS-TV reduces performance degradation on unseen speakers and outperforms all baseline models in multiple datasets. The audio samples are available at \url{https://gzs-tv.github.io/}.
\end{abstract}
\noindent\textbf{Index Terms}: text-to-speech, voice conversion, zero-shot

\section{Introduction}

With the development of speech synthesis methods, applications based on speech synthesis have become increasingly prevalent \cite{DBLP:journals/corr/abs-2106-15561}. Currently, most speech synthesis models learn to synthesize speech for speakers within the training dataset by memorizing their unique rhythm (speaking rate) and timbre (voice characteristics) during training \cite{DBLP:conf/icassp/ShenPWSJYCZWRSA18, DBLP:conf/iclr/0006H0QZZL21, DBLP:conf/nips/GibianskyADMPPR17, DBLP:conf/icml/PopovVGSK21, DBLP:conf/icml/KimKS21}. For example, VITS \cite{DBLP:conf/icml/KimKS21} can synthesize high-quality speech for in-dataset speakers by utilizing the uncertainty modeling over latent variables and adversarial training. However, these approaches cannot synthesize speech for unseen speakers, hence are unable to meet the increasing demand for personalized speech synthesis.

Recent research has proposed speaker adaptive speech synthesis to address this problem. These approaches can be divided into two categories: few-shot speaker adaptation based and zero-shot speaker adaptation based. Few-shot speaker adaptation approaches \cite{DBLP:conf/ssw/TamuraMTK98, DBLP:conf/icassp/Yan0LQZSL21, DBLP:conf/interspeech/XinSTKS20, DBLP:conf/icassp/WangFYTWQW21, DBLP:conf/interspeech/ChoiHKH20, DBLP:conf/icassp/0028HWGLG20} typically pre-train a speech synthesis model on multi-speaker datasets, then fine-tune it with a few speech samples from the unseen speaker. Although these approaches can achieve a good quality of synthesized speech, as demonstrated in \cite{DBLP:conf/icip/WangSJ22, DBLP:conf/iclr/Chen0LLQZL21, DBLP:conf/nips/ArikCPPZ18, DBLP:conf/interspeech/Wu00HZSQL22}, the requirements of computational resources and transcriptions for fine-tuning limit their application scenarios. In contrast, zero-shot speaker adaptation based approaches \cite{DBLP:conf/interspeech/TjandraPZK21, DBLP:conf/icassp/ChienLHHL21, DBLP:conf/icassp/CooperLYFWCY20, DBLP:conf/icassp/ZhangPHL19, DBLP:conf/nips/JiaZWWSRCNPLW18} jointly train a speaker encoder with a speech synthesis model. For instance, the VITS-based zero-shot approach YourTTS \cite{DBLP:conf/icml/CasanovaWSJGP22}, utilizes a speaker encoder to extract speaker embeddings from reference speech and then utilizes these embeddings as input for YourTTS's speech synthesis model to generate speech for unseen speakers. Similarly, StyleSpeech \cite{DBLP:conf/icml/MinLYH21} uses the same method to achieve zero-shot adaptation, and Meta-StyleSpeech \cite{DBLP:conf/icml/MinLYH21} introduces meta-learning for StyleSpeech to improve the quality of synthesized speech. These approaches offer a broad range of application scenarios. However, the limited capability of speaker embedding and distribution shift between seen and unseen speakers present challenges for the models' generalizability, resulting in performance gaps between seen and unseen speakers and poor model performance in zero-shot scenarios \cite{DBLP:conf/icassp/MathurKBL20,DBLP:conf/interspeech/ChoiHKH20}.

\begin{figure*}[ht]
\centering 
\setlength{\abovecaptionskip}{0.1cm}
\subfigure[Training procedure.]{
\label{Fig.sub.1}
\includegraphics[width=0.62\textwidth]{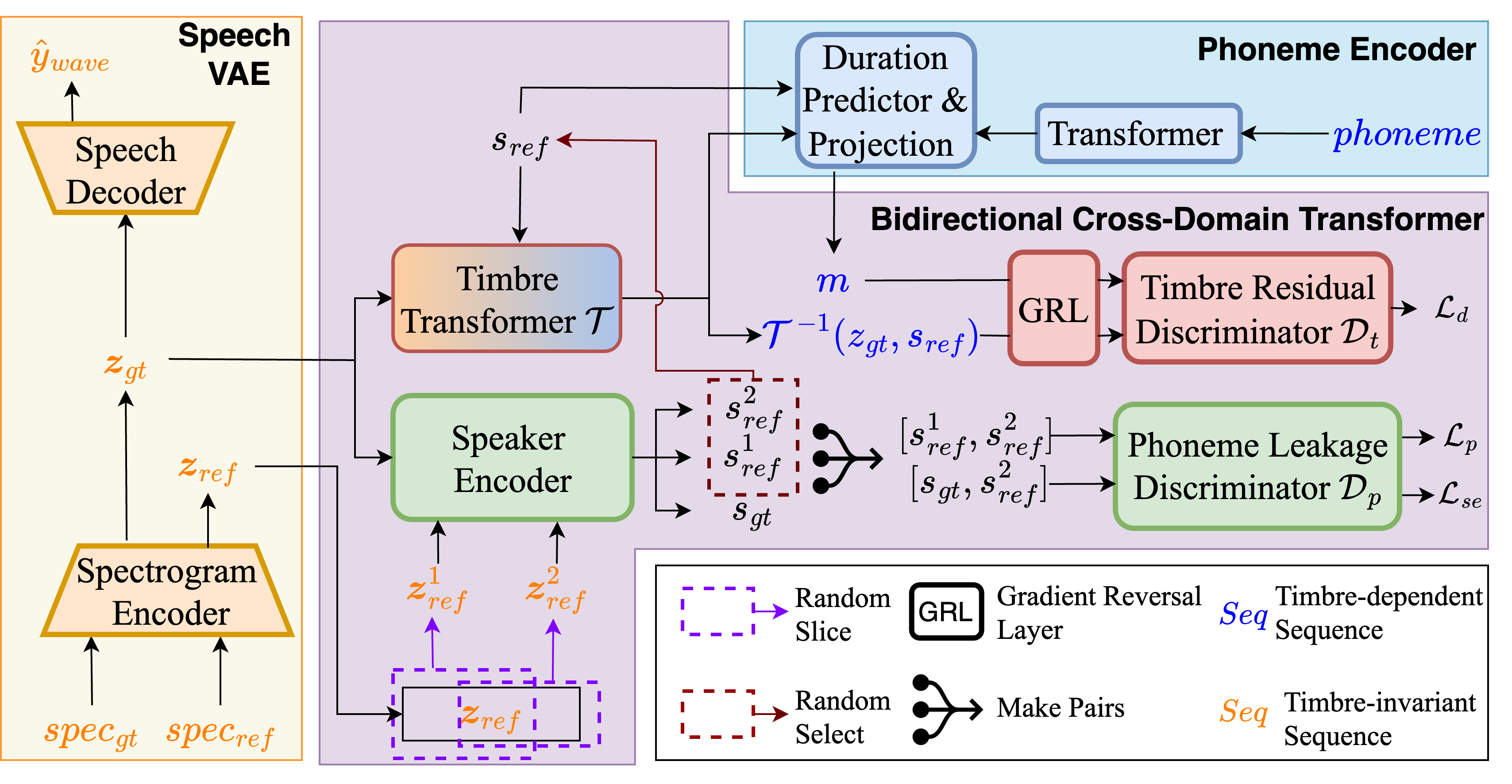}}
\hspace{1mm}
\subfigure[TTS inference.]{
\label{Fig.sub.2}
\includegraphics[width=0.125\textwidth]{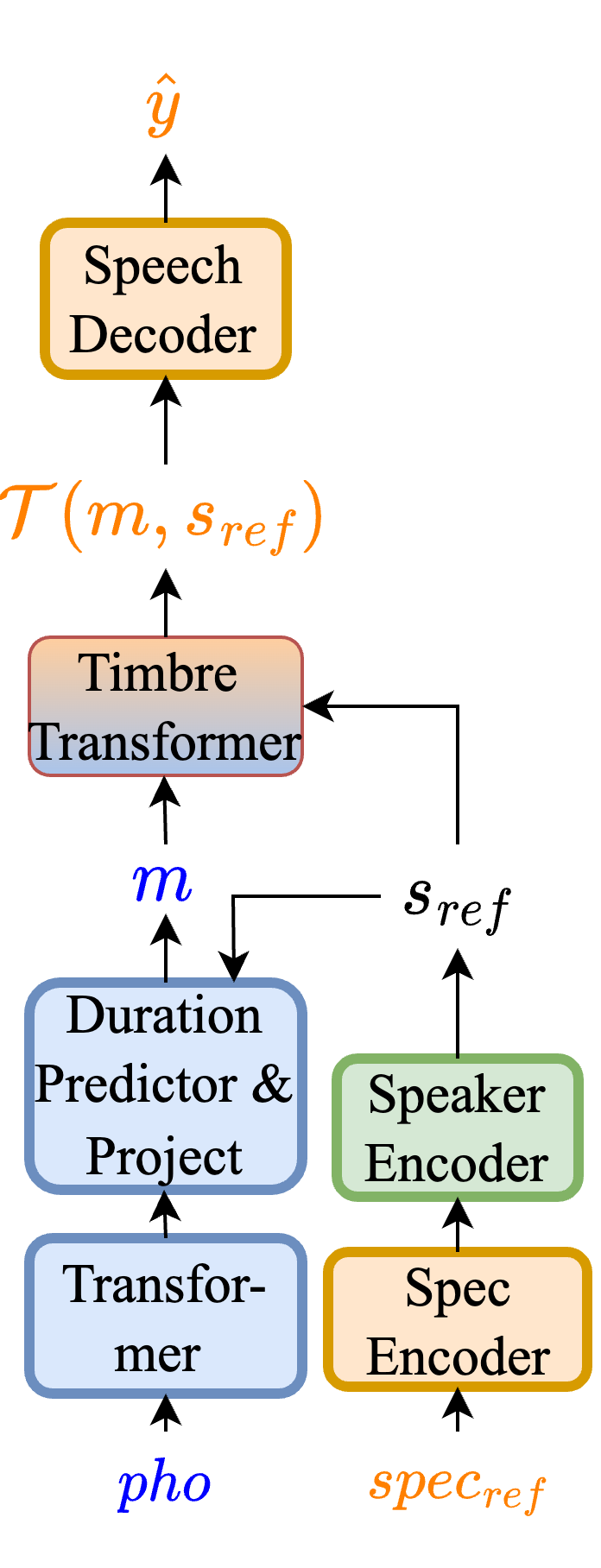}}
\hspace{1mm}
\subfigure[VC inference.]{
\label{Fig.sub.3}
\includegraphics[width=0.14\textwidth]{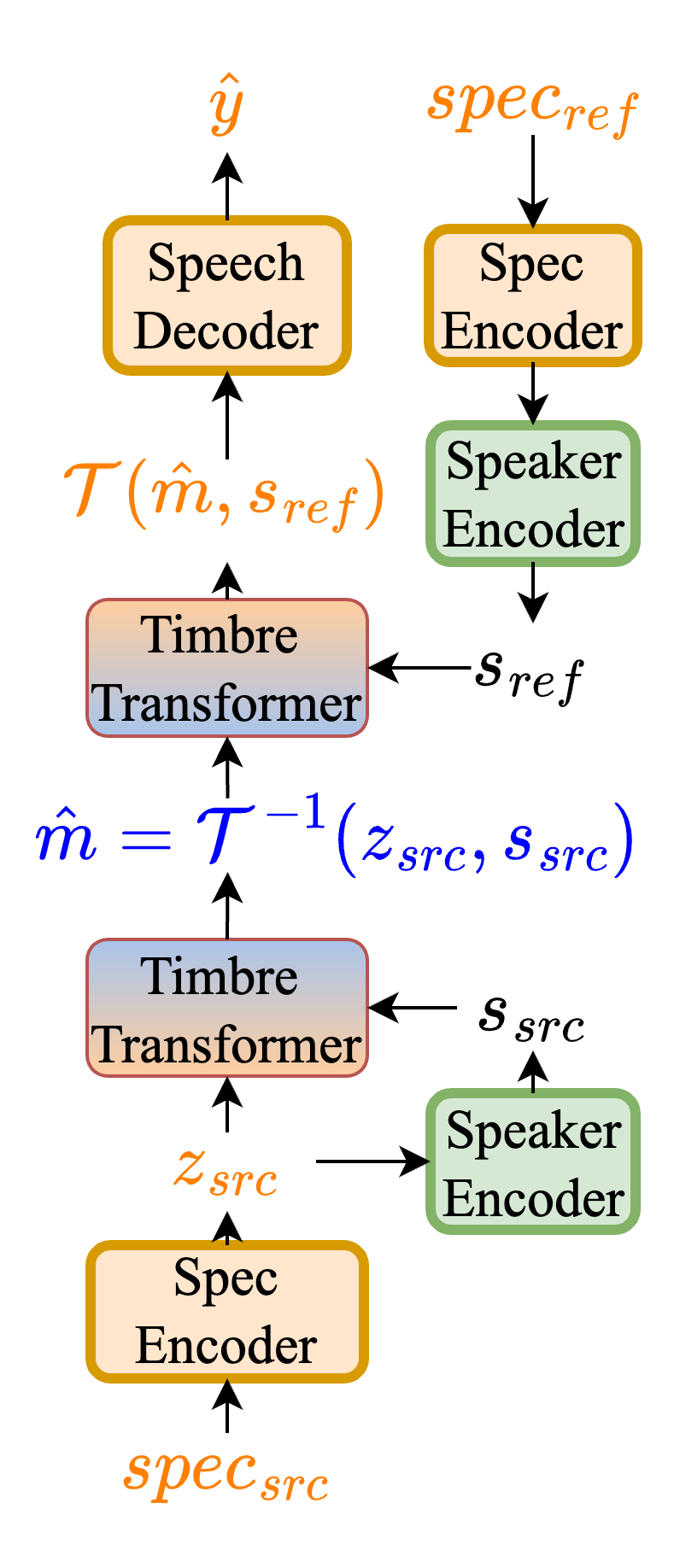}}
\caption{Model diagram depicting (a) training procedure, (b) TTS inference procedure and (c) VC inference procedure.}
\label{Fig.main}
\vspace{-0.2cm}
\end{figure*}

To tackle these issues, inspired by recent research on disentangled representation learning and out-of-distribution generalization problems \cite{DBLP:conf/icml/YiHSSJLM21, DBLP:conf/ijcai/0001LLOQ21, DBLP:conf/nips/LeeKCL21}, we propose GZS-TV, a generalizable zero-shot speaker adaptation text-to-speech (TTS) and voice conversion (VC) model. Since speech contains highly entangled phoneme and timbre information, we aim to disentangle these two types of information to enhance the generalizability of zero-shot speaker-adaptive speech synthesis. Specifically, GZS-TV employs disentangled learning for both speaker information extraction and timbre transformation to avoid phoneme information leakage into the speaker embedding and to better align the timbre of synthesized speech with the ground truth (GT) speech. Additionally, we leverage the representation learning capability of the variational autoencoder (VAE) \cite{DBLP:journals/corr/KingmaW13} to enhance the capacity of the speaker encoder and enrich speaker embedding. The TTS and VC experiments conducted on LibriTTS \cite{DBLP:conf/interspeech/ZenDCZWJCW19} and VCTK \cite{VCTK} datasets demonstrate that GZS-TV is able to synthesize speech that is more natural and more similar to reference speaker's voice than recent state-of-the-art methods.

\section{Methods}

The training, TTS, and VC inference procedures of our GZS-TV model are depicted in Figure 1. The subscripts \(gt\), \(ref\) and \(src\) indicate that the corresponding sequences or embeddings come from the GT, reference, and source speech, respectively. The GT speech is used only during training and is obtained from the seen speakers on the training set, providing phoneme-speech pairs for the model training procedure. The reference speech is also obtained from the same speakers as the GT speech during training, but it can come from any speaker (seen or unseen) during TTS and VC inference procedures. It provides timbre and rhythm information to guide speech synthesis, thereby achieving zero-shot speaker adaptation. The source speech is used only during VC inference and can be from any speaker, providing phoneme and rhythm information for the VC procedure. 

As shown in Figure 1a, GZS-TV comprises three parts: a speech VAE, a phoneme encoder and a bidirectional cross-domain transformer. During training, the \textit{speech VAE} samples frame-level speech representations \(z_{gt}\) and \(z_{ref}\) from linear-spectrogram \(spec_{gt}\) and \(spec_{ref}\), which contain highly entangled phoneme, timbre and rhythm information. It then reconstructs waveform audio \(\hat{y}_{wave}\) from \(z_{gt}\). The \textit{phoneme encoder} encodes and projects the phoneme to a frame-level (same as \(z_{gt}\)) timbre-invariant phoneme representation \(m\). The \textit{bidirectional cross-domain transformer} bridges the timbre-dependent domain (where \(z_{gt}\) belongs) and timbre-invariant domain (where \(m\) belongs) with disentangled representations learning, which is the main component in our model. Specifically, the \textit{speaker encoder} can extract speaker embeddings that represent the speaker's timbre and rhythm characteristics from frame-level speech representations, i.e., extracting speaker embedding \(s_{gt}\) from \(z_{gt}\). The \textit{timbre transformer} can perform lossless bidirectional transformations. The forward transformation fuses the speaker embedding with the timbre-invariant phoneme representation to synthesize a timbre-dependent speech representation, e.g., transforming \(m\) into \(\mathcal{T}(m, s_{ref})\) in Figure 1b. This transformation is used exclusively by the TTS and VC inference procedures. The reverse transformation disentangles and removes the timbre information from the timbre-dependent speech representation based on the speaker embedding and produces a timbre-invariant phoneme representation, e.g., transforming \(z_{gt}\) into \(\mathcal{T}^{-1}(z_{gt}, s_{ref})\) in Figure 1a. This transformation is only used during training and VC inference procedures. Subsequently, two discriminators are designed to improve the model’s generalizability on unseen speakers. Details of the speaker encoder and phoneme leakage discriminator can be found in Sections 2.1 and 2.2. The timbre transformer and timbre residual discriminator are described in Section 2.3. Our GZS-TV is developed on the VITS baseline, which is similar to YourTTS. In contrast to YourTTS, we remove the requirement for speaker embedding as an input for the speech VAE, so that we can extract the speaker embedding from the latent speech representation. Furthermore, design two discriminators to enhance the model’s generalizability to unseen speakers.

During TTS inference procedure (Figure 1b), GZS-TV begins by extracting the speaker embedding \(s_{ref}\) from the reference speech. It then encodes the phoneme sequence $pho$ to a timbre-invariant frame-level phoneme representation $m$, and transforms it into a timbre-dependent speech representation $\mathcal{T}(m, s_{ref})$. Finally, GZS-TV generates the synthesized speech waveform $\hat{y}$ from this speech representation. During VC inference procedure (Figure 1c), GZS-TV first extracts speaker embeddings $s_{src}$ and $s_{ref}$ from the source and reference speech, respectively. It then transforms the spectrogram $spec_{src}$ to a timbre-invariant phoneme representation $\hat{m}$ conditioned on $s_{src}$, and further transforms $\hat{m}$ into a timbre-dependent speech representation $\mathcal{T}(\hat{m}, s_{ref})$. Finally, GZS-TV generates the synthesized speech waveform $\hat{y}$ from this speech representation.

\subsection{Extracting Speaker Embedding from Latent Speech Representation}

A critical challenge in zero-shot speaker adaptation is extracting accurate and generalizable speaker embeddings from a short segment of reference speech. We find that extracting speaker embeddings from high-level speech representations can address this challenge and enhance the generalization of the extracted speaker embeddings, compared to extracting them from raw audio or spectrogram. This improvement can be attributed to two factors. Firstly, the sampling process is a form of data augmentation that adds noise, which can enhance the model's generalizability, as shown in the previous study \cite{DBLP:conf/icml/YiHSSJLM21}. Secondly, the spectrogram encoder's representation learning capability can simplify the distribution of speaker embeddings and enable the speaker encoder to learn richer embeddings, leading to better generalization for unseen speakers. 

In practice, we built the speaker encoder upon ECAPA-TDNN \cite{DBLP:conf/interspeech/DesplanquesTD20}, a popular speaker verification model. To obtain the speaker embedding, we replaced the classifier layers of ECAPA-TDNN with feedforward layers. During each training step of GZS-TV, a speech sample is randomly selected from the same speaker as the GT speech and used as the reference speech. The spectrogram encoder then samples speech representation $z_{ref}$ from this reference speech and divides it into two sub-sequences, $z^1_{ref}$ and $z^2_{ref}$, with an overlap of $\lambda_{ol}$ frames, where $\lambda_{ol}$ is a hyperparameter for the phoneme leakage discriminator. Subsequently, the speaker encoder extracts two speaker embeddings $s^1_{ref}$ and $s^2_{ref}$ from $z^1_{ref}$ and $z^2_{ref}$, respectively. Finally, one of the speaker embeddings is randomly selected as input for downstream modules to provide the speaker's timbre and rhythm information. However, both embeddings are used as input for the phoneme leakage discriminator. More details are available in Section 2.2.

\subsection{Disentangled Representation Learning for The Speaker Encoder to Avoid Information Leakage}

To ensure that the speaker embedding remains free from any leaked speaker-irrelevant information, particularly phoneme information, which could hinder the generalizability of the encoder on unseen speakers \cite{DBLP:conf/ijcai/0001LLOQ21}, we propose a \textit{phoneme leakage discriminator}. This discriminator can detect leakage phoneme information and improve the phoneme-speaker information disentanglement ability of the speaker encoder.

During the training phase, the speaker encoder extracts an additional speaker embedding $s_{gt}$ from the GT speech, in addition to the $s^1_{ref}$ and $s^2_{ref}$ embeddings extracted as described in Section 2.1, as shown in Figure 1a. From these three speaker embeddings, two contrastive embedding pairs are constructed: $[s^{1}_{ref}, s^{2}_{ref}]$ and $[s_{gt}, s^{2}_{ref}]$. The first pair is extracted from overlapped speech representations and would include overlapping phoneme information if there is a phoneme information leakage issue in the speaker encoder. On the other hand, the second pair is extracted from different speech representations and would include rare to no overlapping phoneme information, regardless of the existence of phoneme information leakage. If a well-trained discriminator cannot distinguish which pair of speaker embeddings contains more leaked information, we can assume that the leakage is negligible and can be ignored. 

To this end, we introduce the phoneme leakage discriminator $\mathcal{D}_p$, which aims to detect phoneme information leakage and applies an adversarial penalty to the speaker encoder. The adversarial penalty \(\mathcal{L}_{se}\) for speaker encoder and the training loss for phoneme leakage discriminator \(\mathcal{L}_{p}\) are defined as follows:
\begin{equation}
\mathcal{L}_{p} = \mathop{\mathbb{E}}_{s^{1,2}_{ref},s_{gt}}(\mathcal{D}_{p}(s_{gt}\oplus s^{2}_{ref})-1)^2+(\mathcal{D}_{p}(s^{1}_{ref}\oplus s^{2}_{ref}))^2
\end{equation}
\begin{equation}
\mathcal{L}_{se} =\lambda_{se} \mathop{\mathbb{E}}_{s^{1,2}_{ref}}(\mathcal{D}_{p}(s^{1}_{ref}\oplus s^{2}_{ref})-1)^2
\end{equation}
where \(\lambda_{se}\) is a parameter that adjusts the weight of \(\mathcal{L}_{se}\) and \(\mathcal{D}_{p}\) refers to a feedforward neural network. By playing this min-max game between the speaker encoder and the phoneme leakage discriminator, the speaker encoder is expected to extract embeddings purely related to the speaker's information.

\subsection{Disentangled Representation Learning for Timbre Transformer to Align Timbre Characteristic}

We further design the \textit{timbre transformer} and the \textit{timbre residual discriminator} to align the timbre characteristics of synthesized and GT speech. The timbre transformer is based on the normalizing flow \cite{DBLP:conf/icml/RezendeM15} in VITS. It comprises multiple affine coupling layers \cite{DBLP:conf/iclr/DinhSB17} and provides bidirectional lossless transformations between timbre-dependent and time-invariant sequences to meet different requirements during the training and inference procedures. Since the timbre transformer's transformation is bidirectional and lossless, we argue that enhancing its ability to disentangle and remove timbre information in speech representation during the reverse transformation is equivalent to improving its ability to align the timbre information of synthesized and GT speech during the forward transformation. Thus we utilize the timbre residual discriminator to detect the presence of residual timbre information in the output of reverse transformation and add a penalty to the timbre transformer to improve its inverse transformation. The timbre residual discriminator mainly consists of multiple Res2Net layers \cite{DBLP:journals/pami/GaoCZZYT21}, an attentive statistics pooling layer \cite{DBLP:conf/interspeech/OkabeKS18} and a classification layer. 

Specifically, for the output sequence of the timbre transformer's reverse transformation, as denoted by \(\mathcal{T}^{-1}(z_{gt}, s_{ref})\) in Figure 1a, and the time-invariant sequence generated from the phoneme, as denoted by \(m\) in Figure 1a, we use the timbre residual discriminator \(\mathcal{D}_{t}\) to determine which of them do not contain timbre information. Then, we introduce the gradient reversal layer (GRL) \cite{DBLP:conf/icml/GaninL15} to invert the gradient. Therefore, if the timbre transformer can fool a well-trained timbre residual discriminator, we can assume that its reverse transformation can disentangle and remove most of the timbre information, and its forward transformation can align the timbre information with GT as well as possible. The timbre transformer and timbre residual discriminator will be optimized in different ways according to \(\mathcal{L}_{d}\):
\begin{equation}
\mathcal{L}_{d} = \mathop{\mathbb{E}}_{m, \mathcal{T}^{-1}(z_{gt}, s_{ref})}(\mathcal{D}_{t}(m)-1)^2+(\mathcal{D}_{t}(\mathcal{T}^{-1}(z_{gt}, s_{ref})))^2
\end{equation}
\begin{equation}
\theta\leftarrow\theta-\epsilon(-\lambda_{d}\frac{\partial\mathcal{L}_{d}}{\partial\theta});\ \nu\leftarrow\nu-\epsilon(\frac{\partial\mathcal{L}_{d}}{\partial\nu})
\end{equation}
where \(\theta\) is the parameter of the timbre transformer, \(\nu\) is the parameter of the timbre residual discriminator, \(\epsilon\) is the learning rate, \(\lambda_{d}\) is a weight hyperparameter and \(\mathcal{T}\) is timbre transformer.

\section{Experiments}

\subsection{Experimental Setup}

\textbf{Training Details.} We trained GZS-TV on the clean set of LibriTTS and downsampled all audio samples to 22050 Hz. All modules were trained together in a single training stage. During training, we set both $\lambda_{se}$ and $\lambda_{d}$ to 8, and limited $\lambda_{ol}$ to a range of 20\% to 40\% to avoid the discriminator overwhelming the timbre transformer. We used a batch size of 64, employed the AdamW optimizer \cite{DBLP:conf/iclr/LoshchilovH19}, with $\beta_1=0.8$, $\beta_2=0.99$, and set the weight decay to 0.01. Additionally, we initialized the learning rate to $2\times10^{-4}$, with a decay factor of $\gamma=0.999875$.

\noindent\textbf{Evaluation metrics.} We employ the mean opinion score (MOS) \cite{DBLP:conf/iclr/Chen0LLQZL21} to evaluate the naturalness of synthetic speech. To evaluate the speaker similarity of synthetic speech, we use both the speaker embedding cosine similarity (SMCS) \cite{DBLP:conf/icml/CasanovaWSJGP22} and similarity mean opinion score (SMOS). Both MOS and SMOS are rated on a 1-to-5 scale (1 means worst and 5 means best) by 30 native English speakers through the crowdsourcing form, reported with 95\% confidence intervals. Following previous research \cite{DBLP:conf/icml/CasanovaWSJGP22, DBLP:conf/interspeech/LeiYC0022}, the SMCS is computed by Resemblyzer\footnote{https://github.com/resemble-ai/Resemblyzer}, an off-the-shelf tool for computing speaker embedding. A larger SMCS indicates better speaker similarity. We also provide the word error rate (WER) as the intelligibility metric and a smaller WER indicates more explicitly synthesized speech. We adopt a public pre-trained ASR model\footnote{https://huggingface.co/facebook/wav2vec2-large-960h-lv60-self} for speech transcription.

\begin{table*}[h]
\setlength\tabcolsep{3pt}
\caption{Zero-Shot TTS evaluation results.}
\begin{adjustbox}{width=17cm,center}
\begin{threeparttable}
\scriptsize
\begin{tabular}{c|cccc|cccc|cccc}
\toprule
\(\rm Dataset_{speaker}\)     & \multicolumn{4}{c|}{\(\rm VCTK_{unseen}\)}           & \multicolumn{4}{c|}{\(\rm LibriTTS_{unseen}\)}         &\multicolumn{4}{c}{\(\rm LibriTTS_{seen}\)}\\ \midrule
Metric           & SMCS  & SMOS            & MOS             & WER(\%) & SMCS  & SMOS            & MOS             & WER(\%) & SMCS            & SMOS          & MOS      & WER(\%)               \\ \midrule
Ground-Truth     & -     & -               & \(4.38\pm0.07\) & 5.1 & -     & -               & \(4.32\pm0.09\) & 2.3 & -     & -               & \(4.31\pm0.09\)  & 2.4     \\
Reconstruction   & 0.934 & \(4.57\pm0.07\) & \(4.24\pm0.08\) & 7.4 & 0.954 & \(4.59\pm0.08\) & \(4.20\pm0.09\) & 3.0 & 0.972 & \(4.64\pm0.08\) & \(4.22\pm0.09\)  & 2.8           \\ \midrule 
StyleSpeech \cite{DBLP:conf/icml/MinLYH21}      & 0.730 & \(3.19\pm0.08\) & \(3.19\pm0.09\) & 13.2 & 0.762 & \(3.40\pm0.08\) & \(3.25\pm0.09\) & 6.1 & 0.780 & \(3.58\pm0.08\) & \(3.43\pm0.09\)  & \boldmath\(3.4\)         \\
Meta-StyleSpeech \cite{DBLP:conf/icml/MinLYH21} & 0.738 & \(3.21\pm0.08\) & \(3.18\pm0.09\) & 13.4 & 0.762 & \(3.41\pm0.08\) & \(3.21\pm0.09\) & 6.3 & 0.781 & \(3.59\pm0.08\) & \(3.42\pm0.09\)  & 3.5         \\
YourTTS\tnote{1} \cite{DBLP:conf/icml/CasanovaWSJGP22} & 0.783 & \(3.56\pm0.07\) & \(3.64\pm0.08\) & 11.9 & 0.788 & \(3.62\pm0.06\) & \(3.65\pm0.08\) & 6.0 & 0.806 & \(3.80\pm0.07\)& \(3.78\pm0.06\)   & 5.8 \\
GZS-TV           & \textbf{0.796} & \boldmath\(3.70\pm0.05\) & \boldmath\(3.75\pm0.07\) & \boldmath\(11.2\) & \textbf{0.814} & \boldmath\(3.87\pm0.06\) & \boldmath\(3.77\pm0.07\) & \boldmath\(5.9\) & \textbf{0.825} & \boldmath\(3.98\pm0.08\)& \boldmath\(3.86\pm0.06\) & 5.8  \\ \bottomrule
\end{tabular}
 \begin{tablenotes}
        \scriptsize
        \item 1. Since YourTTS is trained on volume-normalized GT speech, its results are computed on this type of speech too, in the same way as in its original experiment.
      \end{tablenotes}
\end{threeparttable}
\end{adjustbox}
\vspace{-0.3cm}
\end{table*}

\noindent\textbf{Baselines approaches.} We compare GZS-TV with several baselines. 1) Ground-Truth: GT speech. 2) Reconstruction: speech reconstructed from \(z_{gt}\) through speech decoder. 3) StyleSpeech \cite{DBLP:conf/icml/MinLYH21}: a speaker adaptive TTS approach based on style-adaptive layer normalization \cite{DBLP:conf/icml/MinLYH21}. 4) Meta-StyleSpeech \cite{DBLP:conf/icml/MinLYH21}: another version of StyleSpeech based on meta-learning. Both baselines 3 and 4 are trained on LibriTTS. 5) YourTTS \cite{DBLP:conf/icml/CasanovaWSJGP22}: current state-of-the-art zero-shot speaker adaptive TTS model in English, which is based on VITS like ours. By comparing our model with YourTTS, we can demonstrate that our advancement does not solely result from using a more sophisticated speech synthesis backbone. We use their public checkpoint, which is pre-trained on VCTK and fine-tuned on LibriTTS. Since baselines 3, 4 and 5 can only synthesize 16 kHz speech, we downsample the synthesized speech of GZS-TV to 16 kHz when evaluating.

\noindent\textbf{Evaluation datasets.} We utilized 37 out of 39 speakers in the LibriTTS test set (two speakers had insufficient samples) and 108 out of 109 speakers in the VCTK dataset (one speaker lost transcriptions) for the unseen speakers' TTS and VC evaluations. Moreover, we selected 37 random speakers from the LibriTTS training set to evaluate seen speakers. To ensure the diversity of the data, we randomly chose 5 test sentences for each LibriTTS speaker and 2 test sentences for each VCTK speaker. Additionally, we conducted YourTTS's TTS experiment on \mbox{LibriTTS} for reference. Notably, the reference speech in this experiment is considerably longer than in other zero-shot speech synthesis experiments \cite{DBLP:conf/interspeech/CongYXS21, DBLP:conf/interspeech/LeiYC0022}, leading to significantly better outcomes. Furthermore, we examined the impact of reference speech length using SMCS on LibriTTS's unseen speakers. The results are presented in Figure 2.

\begin{figure}[t]
\centering 
\label{SMCS.2}
\includegraphics[width=0.40\textwidth]{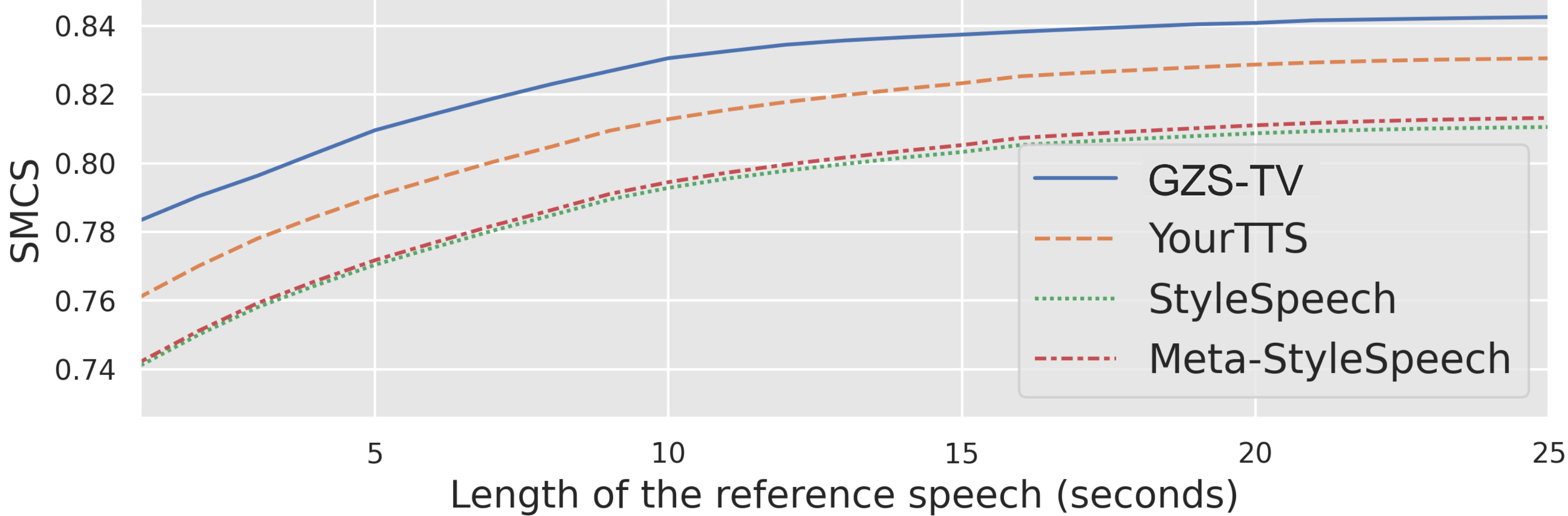}
\caption{Effect of reference speech length on SMCS.}
\label{Fig.SMCS}
\vspace{-0.4cm}
\end{figure}

\subsection{Experimental Results}

\begin{table}[ht]
\vspace{-0.2cm}
\caption{Zero-Shot VC evaluation results.}

\begin{adjustbox}{width=8cm,center}
\scriptsize
\begin{tabular}{c|cc|cc}
\toprule
Dataset    & \multicolumn{2}{c|}{\(\rm VCTK_{unseen \rightarrow unseen}\)}           & \multicolumn{2}{c}{\(\rm LibriTTS_{unseen \rightarrow unseen}\)} \\ \midrule
Metric  & SMOS      & MOS      & SMOS      & MOS       \\ \midrule
YourTTS & \(3.61\pm0.07\)  & \(3.67\pm0.07\) & \(3.68\pm0.06\)  & \(3.68\pm0.07\)\\
GZS-TV  & \boldmath\(3.74\pm0.09\)  & \boldmath\(3.77\pm0.05\)  & \boldmath\(3.91\pm0.04\)  & \boldmath\(3.79\pm0.06\) \\ \bottomrule
\end{tabular}

\end{adjustbox}
\vspace{-0.3cm}
\end{table}

\begin{table}[ht]
\caption{YourTTS's experimental results on LibriTTS.}
\begin{adjustbox}{width=7cm,center}
\scriptsize
\begin{tabular}{c|ccc}
\toprule
Metric  & SMCS      & SMOS      & MOS       \\ \midrule
YourTTS & 0.856 & \(4.17\pm0.08\)  & \(3.79\pm0.10\) \\
GZS-TV & \textbf{0.885} & \boldmath\(4.31\pm0.07\)  & \boldmath\(3.90\pm0.09\)  \\ \bottomrule
\end{tabular}
\end{adjustbox}

\end{table}

The evaluation results in Tables 1 and 2 demonstrate that GZS-TV outperforms StyleSpeech, Meta-StyleSpeech, and YourTTS in almost all metrics. Firstly, GZS-TV exhibits good generalizability by effectively reducing the performance gaps between seen and unseen speakers. These results suggest that GZS-TV can handle out-of-dataset speakers better than other methods. It is noteworthy that YourTTS performs well on VCTK due to its pre-training on this dataset. However, despite using the same speech synthesis backbone as YourTTS, GZS-TV outperforms YourTTS regarding speaker similarity on both seen and unseen speakers, thanks to its disentanglement learning of speaker encoder and timbre transformer. Additionally, cross-dataset speaker adaptation remains challenging, as evidenced by the significant disparities in results on the unseen speakers of LibriTTS and VCTK datasets for all approaches.

\begin{table}[h]

\caption{TTS ablation studies on \(LibriTTS_{unseen}\).}
\begin{adjustbox}{width=8cm,center}
\scriptsize
\begin{tabular}{c|c|ccc}
\toprule
 &Setting  & SMCS      & CSMOS      & CMOS       \\ \midrule
&GZS-TV & \(\rm0.814\) & \(\rm0\)  & \(\rm0\) \\ \midrule
\#1&w/o \(\mathcal{L}_{d}\) & \(\rm0.799\) & \(\rm-0.13\)  & \(\rm-0.01\) \\
\#2&w/o \(\mathcal{L}_{se}\) & \(\rm0.801\) & \(\rm-0.12\)  & \(\rm-0.11\)  \\
\#3&extract \(s\) from \(spec\) & \(\rm0.807\) & \(\rm-0.07\)  & \(\rm-0.05\)  \\ \bottomrule 
\end{tabular}
\end{adjustbox}
\vspace{-0.4cm}
\end{table}

\begin{figure}[h]
\centering 
\setlength{\abovecaptionskip}{0.1cm}
\subfigure[w/o \#2 and \#3.]{
\label{Fig.se.1}
\includegraphics[width=0.18\textwidth]{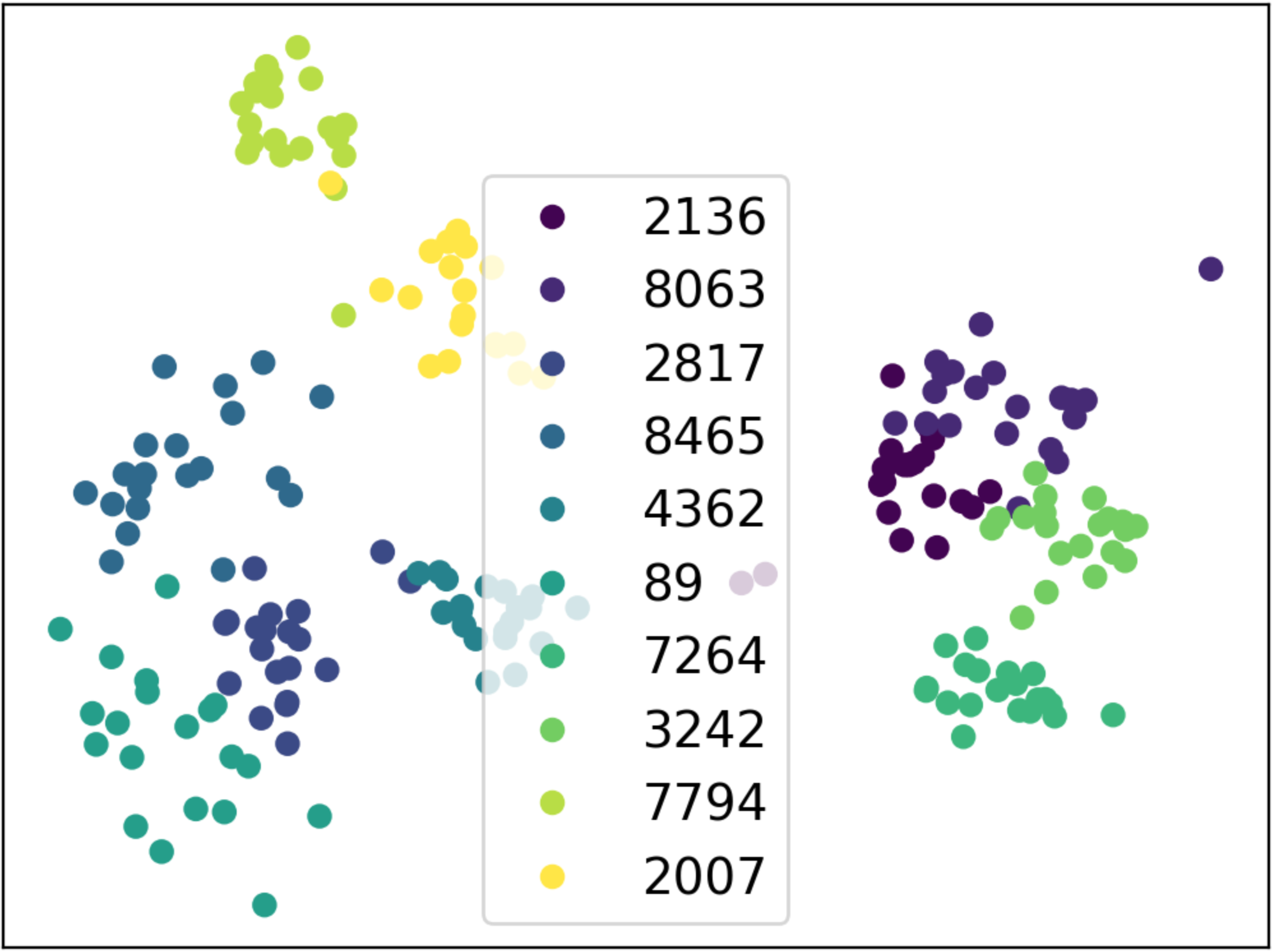}}
\hspace{3mm}
\subfigure[with \#2 and \#3.]{
\label{Fig.se.2}
\includegraphics[width=0.18\textwidth]{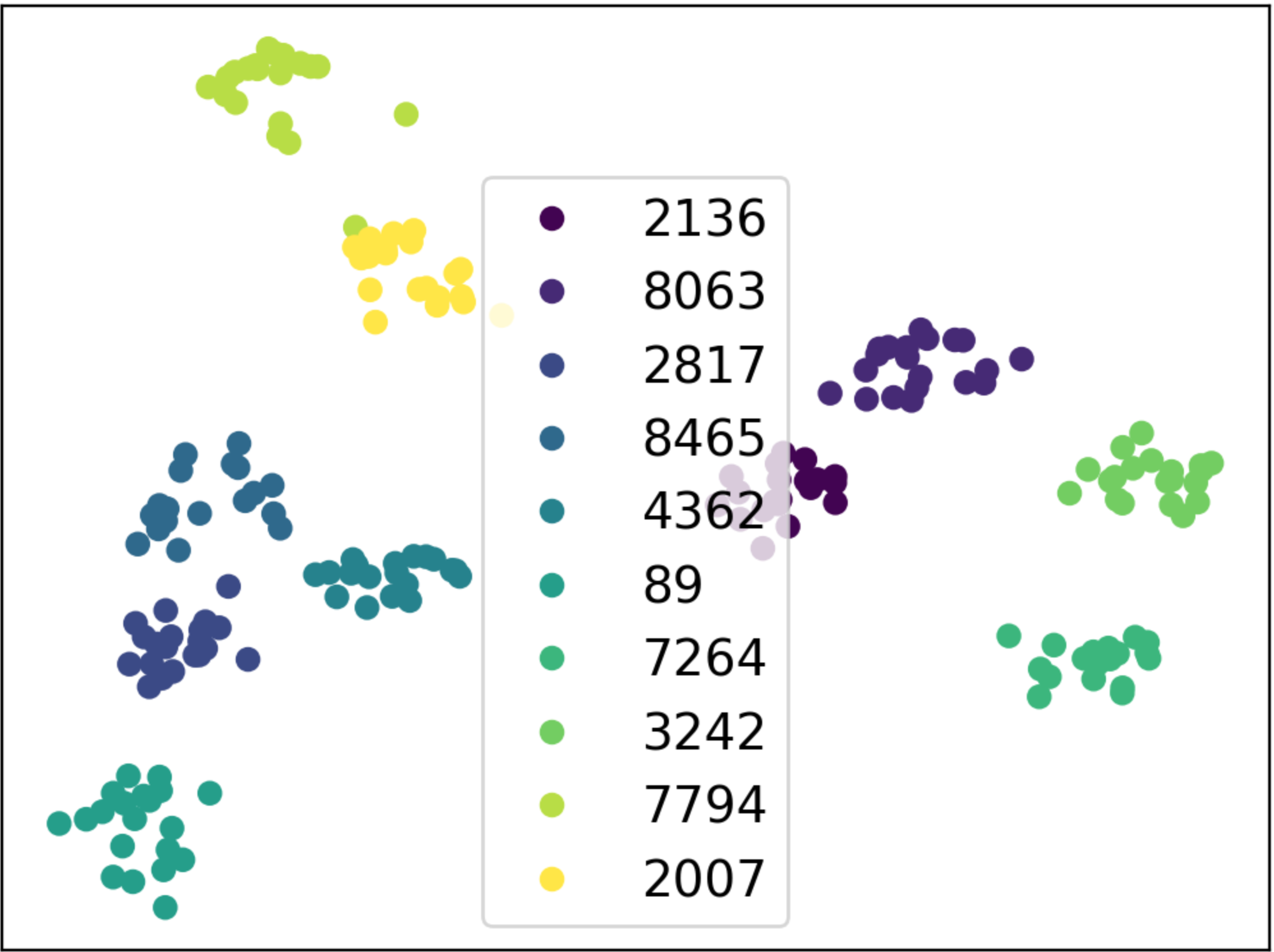}}
\caption{Visualization of speaker encoder embeddings after PCA dimensionality reduction.}
\label{Fig.se}
\vspace{-0.1cm}
\end{figure}

To verify the effectiveness of each module, we conducted ablation studies. As shown in Table 4, removing $\mathcal{L}{d}$ (\#1) led to a drop in speaker similarity, while removing $\mathcal{L}{se}$ (\#2) and using direct extraction (\#3) resulted in reduced speaker similarity and naturalness. We also conducted an ablation study on the VCTK dataset to investigate the effect of (\#2) and (\#3) on the speaker encoder. The visualization results in Figure 3 demonstrate that both (\#2) and (\#3) enhance the generalizability of the speaker encoder, reducing confusion and outliers when extracting embeddings for previously unseen speakers.

\section{Conclusions}
We proposed GZS-TV, a generalizable zero-shot speaker adaptive TTS and VC model. GZS-TV incorporates feature disentanglement in both the speaker encoder and timbre transformer to enhance their generalizability. Additionally, it leverages the representation learning capability of VAE to improve the speaker encoder. Our experiments demonstrate that GZS-TV can synthesize high-quality speech in zero-shot scenarios and significantly reduce the quality gap between seen and unseen speakers. For future research, we plan to explore more effective timbre transformation techniques and investigate strategies to mitigate mispronunciations of TTS in zero-shot scenarios.

\newpage
\pagebreak

\bibliographystyle{IEEEtran}

{\scriptsize
\bibliography{mybib}}

\end{document}